\documentclass[aps,pre,a4paper,preprint,groupedaddress,
showpacs,showkeys,preprintnumbers,floatfix,dvips]{revtex4}

\usepackage[utf8]{inputenc}
\usepackage[T1]{fontenc}
\usepackage{times}
\usepackage{graphicx}
\usepackage{psfrag}
\usepackage{color}
\usepackage{amsmath}
\usepackage{amscd}
\usepackage{amssymb}
\usepackage[section]{placeins}
\usepackage[normalem]{ulem}
\usepackage{caption}
\usepackage{ragged2e}
\usepackage{subfigure}
\usepackage{nicefrac}

\DeclareCaptionJustification{reallyjustified}{\justifying}
\begin{document}

\title{A Generalized Sznajd Model}



\author{Andr\'e M. Timpanaro}
\email[]{timpa@if.usp.br}
\author{Carmen P. C. Prado}
\email[]{prado@if.usp.br}

\thanks{acknowledgements - The authors thank FAPESP for financial support}
\affiliation{Instituto de F\'{i}sica, Universidade de S\~{a}o Paulo \\
  Caixa Postal 66318, 05314-970 - S\~{a}o Paulo - S\~{a}o Paulo - Brazil}


\date{\today}

\begin{abstract}

 In the last decade the  Sznajd Model has been successfully 
employed in modeling some  properties and scale
 features of both proportional and majority elections. 
We propose a new version of the Sznajd model with a generalized 
\emph{bounded  confidence} rule - a rule that limits the convincing 
capability  of agents  and that is essential to
allow coexistence of opinions in the stationary state. With an
appropriate choice of parameters it can be reduced to previous models.
We solved this new model both in a mean-field approach (for an
arbitrary number of
 opinions) and numerically in a Barab\'asi-Albert network 
(for three and four opinions), studying the
transient and the possible stationary states. We built the
phase portrait for the special cases of three and four opinions, defining the
attractors and their basins of attraction. Through this analysis, we
 were able to understand and explain discrepancies between mean-field
 and simulation results obtained  in previous works for the usual 
 Sznajd Model with bounded confidence and three opinions. Both the dynamical 
 system approach and our generalized bounded confidence rule are quite general 
 and we think it can be useful to the understanding of other similar models.

\end{abstract}

\pacs{02.50.Ey, 02.60.Cb, 05.45.-a, 89.65.-s, 89.75.-k, 89.75.Hc}

\maketitle

\section{Introduction}

The Sznajd model (SM), proposed in 2000 by Sznajd-Weron and Sznajd, 
is a model that has been successfully employed in the reproduction of some
properties observed in the dynamics of opinion
propagation in a closed community \cite{Sznajd-2000}. It
is a very simple \emph{Ising-like} model, that always leads to a
stationary state of consensus, but with a rich transient behavior. Its
originality resides in the way the state of the sites evolves:
two agreeing sites work together in changing their neighbors' 
state, instead of being influenced by the environment like in
 the voter model \cite{Liggett-1975, Krapivsky-2003}.

This model has been extensively studied, either with its original set
of rules, or in a variety of versions in which one or more rules were
changed in order to describe or include specific features (see for
instance \cite{Stauffer-2003}). It was extended  to higher 
dimensional lattices in 2000 \cite{Stauffer-2000,Bernardes-2001} and
adapted to deal with more than two opinions in networks of
different topologies in 2002 \cite{Bernardes-2002}. Also, in  2002, 
Stauffer \cite{Stauffer-2001} adapted the \emph{bounded confidence} 
restriction, first introduced by Defuant \cite{Defuant-2000} and Krause 
\cite{Krause-2002} in models with continuous opinions, 
to the discrete Sznajd Model scenario. His numerical simulation results, 
for the regular square lattice with three opinions, were in disagreement with 
what was found in 2004 by Schulze \cite{Schulze-2004}, that
solved the same model in a mean-field approach.

In this work we propose a new model, with a distinct  way  
of introducing the idea of limited persuasion among 
electors in a discrete model, as the SM. 
With adequate choices of parameters our model
restores the set of rules of previous works, allowing comparisons.

We solved the model's master equation numerically in a mean-field
approach and 
made simulations of the model in a Barab\'asi-Albert network and 
in a square lattice, looking at the transient behavior, as well as the 
stationary state. With the aid of dynamical systems techniques, 
we were able to draw a general picture of its behavior 
in phase space, identifying its fixed points and basins of attraction, for 
three and four dimensions. This analysis allowed us to understand the
origin of the 
discrepancies in the results obtained by Stauffer \cite{Stauffer-2001} 
and Schulze
\cite{Schulze-2004} for three opinions. We also studied the transient
behavior of the model for three opinions. 
Comparing the mean-field time series with
the ones from the model in a BA network, we see that they share many features, 
which does not happen with the square lattice time series.
 
This paper is organized as follows: In the next section we briefly
review the rules and the behavior of the original SM, 
the new set of rules introduced by Bernardes
and co-authors in 2002, the version of the model with bounded
confidence introduced by Stauffer in 2002, and finally, present 
our own model. In section III we solve the new 
 model in the mean-field approach, in the general case of $M$
 opinions; in section IV we present simulation results, the detailed
 phase-portrait in the special cases of three and four opinions and 
 discuss previous results.  
Finally, in section V, we summarize our conclusions. 

\section{The Sznajd Model}

In the original SM \cite{Sznajd-2000}, 
the sites of a chain with periodic
boundary conditions represented voters, that could only have 
opinions (states) $\sigma=\pm 1$.
If a pair of adjacent sites had the same opinion, they would convince 
their neighbors with probability p=1; however, if they disagreed 
their divergence would be propagated,
with the neighbors adopting an opposite opinion. 
This model always evolves to one of two absorbing states: 
a ferromagnetic state (consensus state), with all voters 
with the same opinion,
or  an  anti-ferromagnetic state, in which every site has an
opinion that is different from the opinion of its neighbors 
(only possible if the chain has an even number
of sites or the periodic boundary is dropped). The transient, however,  
displays a rich behavior, that called the attention of some physicists
\cite{Bernardes-2001}. 

This model has been extensively studied, either with its original set
of rules, or in a variety of versions in which one or more rules were
changed in order to describe or include specific features, as the
possibility of more than two opinions, diffusion of agents,
restrictions in the convincing capability of agents, or different
topologies in the network defining the relationship among voters (for
a review, see for instance \cite{Stauffer-2003}). In most of the
works that followed \cite{Sznajd-2000}, the divergence 
propagation rule was abandoned.

\subsection{The Sznajd model in complex networks }
\label{model}

In \cite{Bernardes-2002}, 
Bernardes \emph{et al.} studied a new version of the SM, that was 
adapted in order to describe the evolution of $N>2$ opinions in
voters located in an arbitrary network.
This new model was employed to simulate proportional elections 
with $M$ candidates in a Barab\'asi-Albert network. In their version, 
each site could be in one of $M + 1$ states, the extra state standing for
undecided voters. Some changes were also introduced in the updating
rules, the idea being that, at each time step the same average number
of neighboring sites were convinced, as in the SM.
This can be accomplished by setting the probability that a 
site convinces another one
to $p = 1/q$, where $q$ is the \emph{degree} (number of
neighbors) of the convincing site (in the SM, $\;p = 1\;$ always).
Also, a different set of rules was devised for the undecided voters, that
were not able to propagate their lack of opinion but could  be
convinced  by one of its decided neighbors, 
even if it did not belong to a pair. 

More precisely, the model is defined by the following set of rules:
Let $\sigma(i,t)$ be the opinion of a site $i$ at time $t$
($\sigma(i,t)\in\{0,1,\dots,M\}$, where the positive values
represent candidates and $\sigma(i,t)=0$ stands for undecided voters). 
Initially, all $N$ voters are undecided, except for a 
set of $M$ \emph{initial electors}  (one
for each candidate), chosen at random.

The dynamics consists in visiting each voter 
in a random (non-sequential) order, applying
the following rules: 

\begin{enumerate}
\item [(I)]A voter $i$ is chosen at random. If it is not 
undecided ($\sigma(i,t)\neq0$), a site $j$ is picked up (at random)
from the set $\Gamma_{i}$ of
neighbors of $i$ and rule II is applied, else nothing happens.
\item [(IIa)]If voter $j$ is undecided ($\sigma(j,t)=0$), then $j$
adopts $i$'s opinion with
probability $p_i = 1/\,q_{i}$, where $q_{i}$ is the 
degree of site $i$.
\item [(IIb)]If both $i$ and $j$ have the same opinion, voter $i$
tries to convince each one of its neighbors with probability
$p_i = \nicefrac{1}{q_{i}}$;
\item [(IIc)]If $i$ and $j$ have different opinions, nothing happens.
\end{enumerate}

\noindent Like the original SM, this model always evolves towards a 
consensus absorbing state, but during the transient this new model 
displays a power-law distribution of candidates with $v$ votes. 
This behavior is in agreement with what has been observed in the 
statistics of real proportional elections 
\cite{Costa-1999, Gonzales-2003, Tese-Fabio}.

\subsection{Bounded Confidence}

In all versions of the SM \footnote{Without the divergence propagation
  rule. In the original version a paramagnetic state was possible.} 
  in which a site $i$ always convinces
another one \emph{independently of its opinion}, the system  evolves
to an absorbing state, in which only one opinion survives. That is not
always the case in real communities. In an attempt to allow the
emergence of different \emph{factions}, 
Defuant et al. \cite{Defuant-2000}, and 
Hegselmann and Krause \cite{Krause-2002} introduced the
idea of \emph{bounded confidence} for models with continuous opinions
($\sigma \in [0,1]$). 
The idea is to assume that 
electors $i$ and $j$, with opinions $\sigma_i$ and $\sigma_j$, can
interact only if $|\sigma_i - \sigma_j|\leq\epsilon$, that is, if
their opinions are close enough. This model also evolves to
absorbing states, but now eventually with the coexistence  of 
two or more opinions that do not interact with each other. This rule
can be easily adapted to discrete models, as the SM, if opinions are
labeled  from 1 to $M$ and $\epsilon$ is set to 1.
 
In 2002 Stauffer studied the SM with bounded confidence in square lattices
 \cite{Stauffer-2001}, while Schulze \cite{Schulze-2004} studied its  mean
field version, arriving at different conclusions: In the lattice,
Stauffer showed that the model `almost always' evolved to an absorbing
state of consensus, while in the mean-field approach presented by
Schulze the consensus was achieved only in 50\% of the cases; 
the system ended up in a state of coexistence of opinions in the
other cases.

\subsection{Generalized bounded confidence rule}

In this work we propose a  new model with a generalization of 
the \emph{bounded confidence}  idea. This new model includes, in a
single set of rules, the original SM, the complex SM, the SM with
bounded confidence proposed by Stauffer, as well as many other
possibilities. 

In our generalized model, a site with opinion $\sigma'$ has a probability 
$p_{\sigma'\rightarrow\sigma}$ of being
convinced by another site with opinion $\sigma$. As in previous
models,  random non-sequential update is employed, and there are 
no undecided voters. 
The rules become:
 
\begin{itemize}
\item[(I$'$)] Choose a voter $i$ at random and a voter
  $j\in\Gamma_{i}$. If  $\sigma_i \neq \sigma_j$ we do nothing,
else we apply rule II$'$.
\item [(II$'$)] Site $j$ tries to convince each one of its neighbors $k$ of
  opinion $\sigma_j=\sigma_i$ with probability  $\displaystyle 
\frac{p_{\sigma_k\rightarrow  \sigma_j}}{q_j}$, 
where $q_j$ is the coordination of site $j$.
\end{itemize}

\noindent Alternatively, rule (II$'$) can also be:
\begin{itemize}
\item[(II$''$)] A neighbor $k$ of $j$ is chosen at random and $j$ 
convinces $k$
with probability $p_{\sigma_k\rightarrow\sigma_j}$.
\end{itemize}

Note that no assumptions about the probabilities 
$p_{\sigma\rightarrow\sigma'}$
are made beforehand. If $p_{\sigma\rightarrow\sigma'}$ is always 
0 or 1, then, with a convenient
choice of  values, one can recover both the usual SM and 
the discrete version with bounded confidence introduced by Stauffer.

\section{Time evolution in a mean-field approach}

We can easily write the master equation for the model presented in the
last section:

\[
\Delta P(\sigma_k = \sigma) = \frac{1}{N} \sum_{\sigma'} 
\sum_{j\in\Gamma_k} \sum_{i\in\Gamma_j} \frac{1}{q_iq_j} 
(p_{\sigma' \rightarrow \sigma} P(\sigma_i = \sigma_j = 
\sigma, \sigma_k = \sigma') - 
\]
\begin{equation}
-p_{\sigma\rightarrow\sigma'} P(\sigma_i = \sigma_j = \sigma',
\sigma_k = \sigma)),
\label{mestra}
\end{equation}

\noindent where $\Gamma_X$ and $q_X$ are, respectively, the set of 
neighbors and the
coordination (degree) of site $X$, and $N$ is the total number of sites. 

If

\[
\eta_{\sigma} = \frac{1}{N}\sum_i P(\sigma_i = \sigma),
\]

\noindent in a mean-field approach, the master equation is reduced to 

\[  
\Delta \eta_{\sigma} = \frac{1}{N}\sum_{\sigma'} 
(\eta_{\sigma}^2 \eta_{\sigma'}p_{\sigma'\rightarrow\sigma} - 
\eta_{\sigma'}^2 \eta_{\sigma}p_{\sigma\rightarrow\sigma'}),
\]

\noindent and in the thermodynamic limit ($N\rightarrow \infty$) we
have

\begin{equation}
\dot{\eta}_{\sigma} = \sum_{\sigma'} (\eta_{\sigma}^2 
\eta_{\sigma'}p_{\sigma'\rightarrow\sigma} - \eta_{\sigma'}^2 
\eta_{\sigma}p_{\sigma\rightarrow\sigma'})\,\,\,\forall\,\,\sigma,
\label{fluxo}
\end{equation}

\noindent where a time-unit corresponds to a Monte Carlo step 
($N$ random trials).

We also define $\vec{\eta} = (\eta_1, \eta_2, \ldots, 
\eta_M)$, $F_{\sigma}(\vec{\eta}) = \dot{\eta}_{\sigma}$ and $\vec{F} = 
(F_1, F_2, \ldots, F_M)$.

As the sum over $\sigma'$ in (\ref{fluxo}) is antisymmetric with 
respect to $\sigma$ and $\sigma'$, we get

\begin{equation}
\sum_{\sigma} \dot{\eta}_{\sigma} = 0 \Rightarrow \sum_{\sigma}
\eta_{\sigma} = \mathrm{constant}.
\label{norm}
\end{equation}

\noindent From the definition of $\eta$ it follows that this 
constant must be equal
to 1. As a consequence, although the flux has $M$ variables,
it is restricted to $M-1$ dimensions. This also implies that zero is an
eigenvalue of the Jacobian of $\vec{F}$ for all values of $\vec{\eta}$.
Also, if $\;\eta_{\sigma} \geq 0 \;\,\forall\,\;\sigma$, the negative
term of $\dot{\eta}_{\sigma}$ in \eqref{fluxo} is proportional to $\eta_{\sigma}$
(and the term multiplying it does not diverge in the limit 
$\eta_{\sigma}\rightarrow 0$). So the flux is
restricted to the region in which all variables are positive (as it
should, since $\eta_{\sigma}$ is the probability that a site chosen at random
has opinion $\sigma$ in the mean-field). 

The region in phase space where $\eta_{\sigma}\geq 0$ and
$\sum_{\sigma}\eta_{\sigma} = 1$ is a
regular simplex, and each vertex $P_i$ represents the consensus
absorbing state of opinion $i$.
The points inside a simplex are unique convex combinations of its
vertices, suggesting a nice way of representing the phase space 
of the problem. The point $P$ that represents the state 
$(\eta_1, \eta_2, \eta_3, \ldots, \eta_M)$ is given by

\[
P = \sum_{\sigma} \eta_{\sigma} P_{\sigma}.
\]

The fixed points of (\ref{fluxo}) are given by

\[
\sum_{\sigma'}\eta_{\sigma'}(\eta_{\sigma}p_{ \sigma' \rightarrow
  \sigma} - \eta_{\sigma'}p_{ \sigma \rightarrow \sigma'}) = 
0\,\,\mathrm{or} \,\,\eta_{\sigma} = 0,
\]

\noindent and the Jacobian matrix of $\vec{F}$, 
$(\mathcal{J}_{\vec{F}})_{\sigma, \sigma'}$ is

\[
\begin{split}
(\mathcal{J}_{\vec{F}})_{\sigma, \sigma'} = &
\frac{\partial F_{\sigma}}{\partial \eta_{\sigma'}} = 
\delta_{\sigma, \sigma'}\sum_{\sigma''}
\eta_{\sigma''}(\eta_{\sigma}p_{ \sigma'' \rightarrow \sigma} - 
\eta_{\sigma''} p_{\sigma \rightarrow \sigma''}) + \\
& +\eta_{\sigma}\left(\delta_{\sigma, \sigma'}\sum_{\sigma''} 
\eta_{\sigma''}p_{ \sigma'' \rightarrow \sigma} + 
\eta_{\sigma}p_{ \sigma' \rightarrow \sigma} - 
2\eta_{\sigma'}p_{\sigma \rightarrow \sigma'}\right).
\end{split}
\]

So, in the fixed point, we have:

\[
(\mathcal{J}^*_{\vec{F}})_{\sigma, \sigma'} = \left\{
\begin{array}{l}
\eta_{\sigma}\delta_{\sigma, \sigma'}\sum_{\sigma''} 
\eta_{\sigma''}p_{ \sigma'' \rightarrow \sigma} + 
\eta^2_{\sigma}p_{ \sigma' \rightarrow \sigma} - 
2\eta_{\sigma}\eta_{\sigma'}p_{\sigma \rightarrow \sigma'} 
\,\,\mathrm{if}\,\,\eta_{\sigma}\neq 0 \\ \\
-\delta_{\sigma, \sigma'}\sum_{\sigma''}\eta^2_{\sigma''}
p_{ \sigma \rightarrow \sigma''}\,\,\mathrm {if}\,\,\eta_{\sigma}=0
\end{array}
\right.
\]

From the expression above it is possible to derive 
the following conclusions:

\begin{itemize}
\item[(a)] Consider first that a fixed point $P^*$ lies in the 
intersection of manifolds of the type $\eta_{\sigma} = 0$, so we have 
(conveniently reordering the variables)

\begin{equation}
\mathcal{J}^* = 
\begin{bmatrix}
\mathcal{J}^*_{R} & \mathcal{M}\\
0 & \mathcal{D}
\end{bmatrix}
\label{eq:jacobiano}
\end{equation}

where $\mathcal{J}^*_{R}$ is the Jacobian restricted to the non-zero 
variables in the fixed point and $\mathcal{D}$ is the Jacobian
restricted to the variables equal to zero in the fixed point, 
which is a diagonal matrix. So for each opinion $\sigma$ such 
that $\eta_{\sigma}^* = 0$ we have an associated eigenvalue 
$\lambda_{\sigma} \leq 0$,

\[
\lambda_{\sigma} = -\sum_{\sigma'}(\eta_{\sigma'}^*)^2p_{\sigma \rightarrow \sigma'}.
\]

\noindent It follows from (\ref{eq:jacobiano}) that if $x$ is an 
eigenvector of $\mathcal{J}^*_R$ with eigenvalue $\lambda$, then

\[
\begin{bmatrix}
\mathcal{J}^*_{R} & \mathcal{M}\\
0 & \mathcal{D}
\end{bmatrix}.
\begin{bmatrix}
x\\
0
\end{bmatrix}=
\begin{bmatrix}
\mathcal{J}^*_{R}.x\\
0
\end{bmatrix}=
\lambda
\begin{bmatrix}
x\\
0
\end{bmatrix}.
\]

So these eigenvectors and eigenvalues are the same as if we had a 
model with fewer opinions. The eigenvectors are also parallel to 
all the manifolds $\eta_{\sigma} = 0$ where $P^*$ is. On the other 
hand, if $\vec{v}^k$ is an eigenvector with
eigenvalue  $\lambda_k$ it follows that
$\mathcal{J}^* \vec{v}^k = \lambda_k \vec{v}^k$. So for 
coordinate $\sigma$ we have $\,v^k_\sigma (\lambda_k - \lambda_{\sigma}) = 0$.

Hence, if $\vec{v}^k$ is not parallel to the manifold defined
by $\eta_{\sigma} = 0$, then we must have 
$\lambda_k = \lambda_{\sigma} \leq 0$.

\item[(b)] We focus now on the possible values that 
$\lambda_{\sigma}$ may take. If

\begin{equation}
\label{cond1}
\eta_{\sigma'}^*p_{\sigma \rightarrow \sigma'} \neq 0,
\end{equation}

\noindent for some $\sigma'$, 
then $\lambda_{\sigma} < 0$ and the flux, in the
neighborhood of $P^*$, is attracted to the manifold $\eta_{\sigma} = 0$. 
Note that the condition expressed in eq. (\ref{cond1})
is equivalent to saying that there are still sites able to convince a 
site with opinion $\sigma$.

On the other hand, if condition (\ref{cond1}) is not satisfied, 
$\lambda_{\sigma} = 0$ and we have, for a point
arbitrarily close to the fixed point (but outside $\eta_{\sigma} = 0$),

\[
\dot{\eta}_{\sigma} \simeq \sum_{\sigma'}
\eta_{\sigma}^2\eta_{\sigma'}^*p_{ \sigma' \rightarrow \sigma},
\]

\noindent which is the time evolution in second order. Therefore, if
there is a $\sigma'$ such that 
$\eta^*_{\sigma'}p_{\sigma' \rightarrow \sigma} \neq 0$, then the manifold
$\eta_{\sigma} = 0$ is unstable in the neighborhood of the fixed
point. If this new condition is also not satisfied, then $\sigma$ 
inevitably interacts only with opinions that don't survive in $P^*$, 
which is the third order term.
 
Let $\Omega$ be the set of opinions $\sigma$ such that
$\eta_{\sigma}=0$ in $P^*$, and let $M$ be the manifold 
$\eta_{\sigma} = 0 \,\forall\,\sigma\in\Omega$, so $P^* \in M$.
It follows that, if any opinion in $\Omega$ interacts in second order, 
then the trajectories are repelled from $M$ in the neighborhood 
of $P^*$. If all of them interact in first order, then the trajectories
are attracted to $M$. If they all interact only in third order the 
model is degenerate, as opinions in $\Omega$
do not interact with opinions outside of it 
(because of the particular choice of $p_{\sigma\rightarrow\sigma'}$).
If all opinions in $\Omega$ either interact in first or 
in third orders the model degenerates asymptotically.

\item[(c)] Suppose now that all probabilities  
$p_{ \sigma \rightarrow \sigma'}$ are non
  zero, meaning that all sites have some chance of convincing any
  other one. Consider a surface formed by moving the boundary of the simplex
  inwardly by a sufficiently small amount (but non zero). With the 
same reasoning   presented above, we conclude that the flux must come from the
  simplex's inner   region, crossing the surface towards the boundary points. 

It follows then that there is an unstable region where all opinions 
coexist. For $M=3$ (3 opinions) this region is a node.

\item[(d)] Finally, consider a manifold in which all opinions do not
  interact with each other, that is, 

\begin{equation}
\nonumber 
\eta_{\sigma},\eta_{\sigma'}\neq 0 \Rightarrow p_{ \sigma' \rightarrow \sigma} = 
p_{ \sigma \rightarrow \sigma'} = 0.
\end{equation}

\noindent Every point in this manifold is a fixed point, and so, 
this manifold may have a basin of
attraction.
\end{itemize}

These arguments give a qualitative idea of the evolution of the
model in the mean-field approach. The less relevant opinions
disappear quickly, and the system has a high probability of
ending up in a state where different and non-interacting opinions
coexist (provided these states exist).

\section{The special cases of 3 and 4 opinions}

In the following sub-sections we will analyze in detail the phase
portraits that represent the dynamics of our model in the cases of three and
four opinions, in which they can be drawn. We will also
present some results about the time evolution of the average 
number of votes for each candidate (the transient behavior) and 
make comparisons
between the mean-field (integrated master equation) and the simulated model
(in BA networks and square lattices).

\subsection{Scenario with three opinions}

It follows from (\ref{norm}) that $\eta_1 + \eta_2 + \eta_3 = 1$,
so that the flux is restricted to an equilateral triangle. A point in
this triangle represents uniquely a set of normalized variables 
$\eta_1, \eta_2, \eta_3$; the
vertices $P_1$, $P_2$ and $P_3$ represent consensus states with
opinions 1, 2 and 3 respectively; and the 
side $A_{i,j}$, connecting  $P_i$ to $P_j$, represents the set of
states in which opinions $i$ and $j$ coexist. 

\begin{figure}[htbp]
\subfigure[Same convincing strength \label{grafico-C}]
{\centerline{\includegraphics[width=8.0cm]{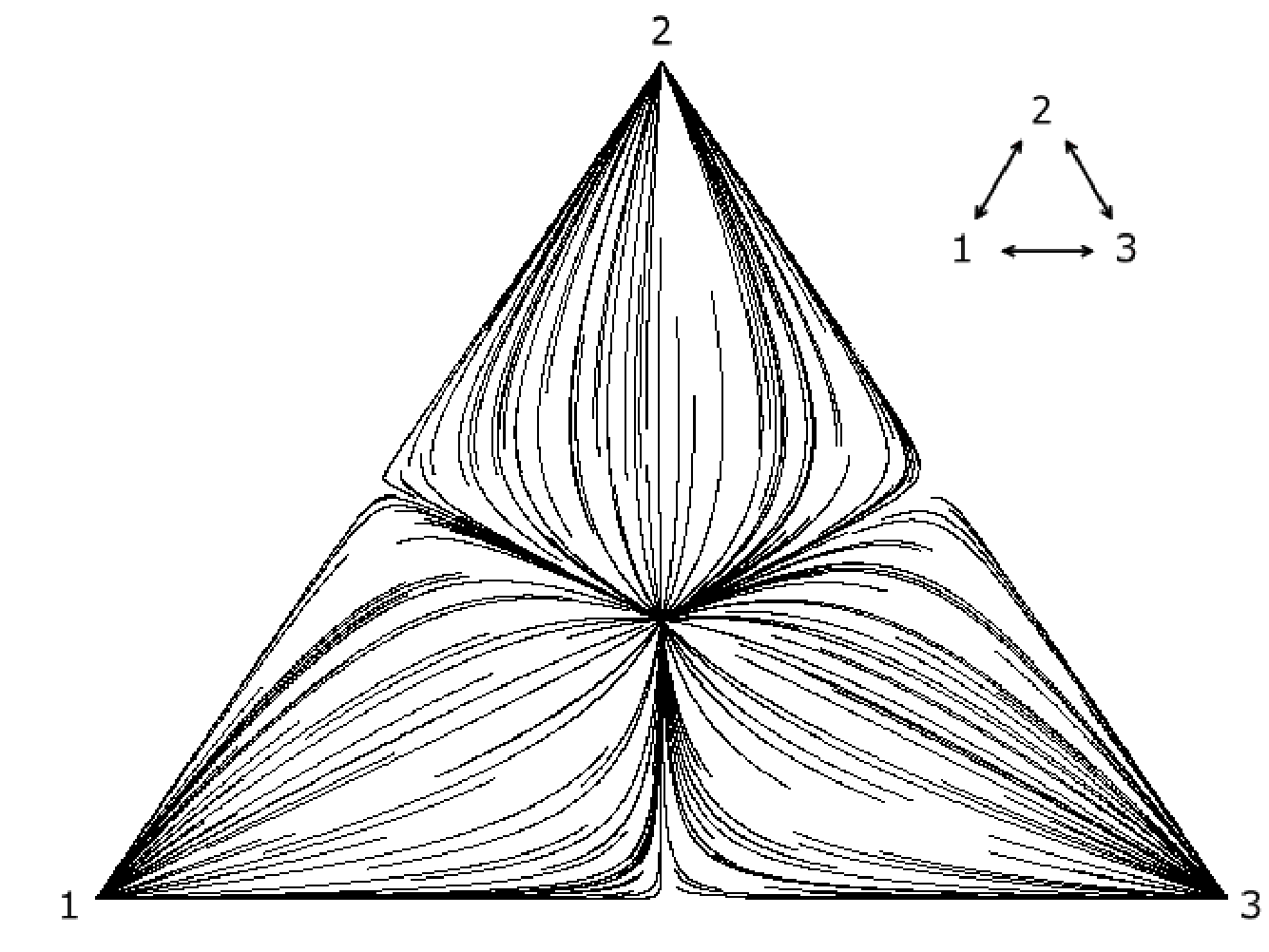}}}
\subfigure[Asymmetric convincing capability \label{grafico-D}]
{\centerline{\includegraphics[width=8.0cm]{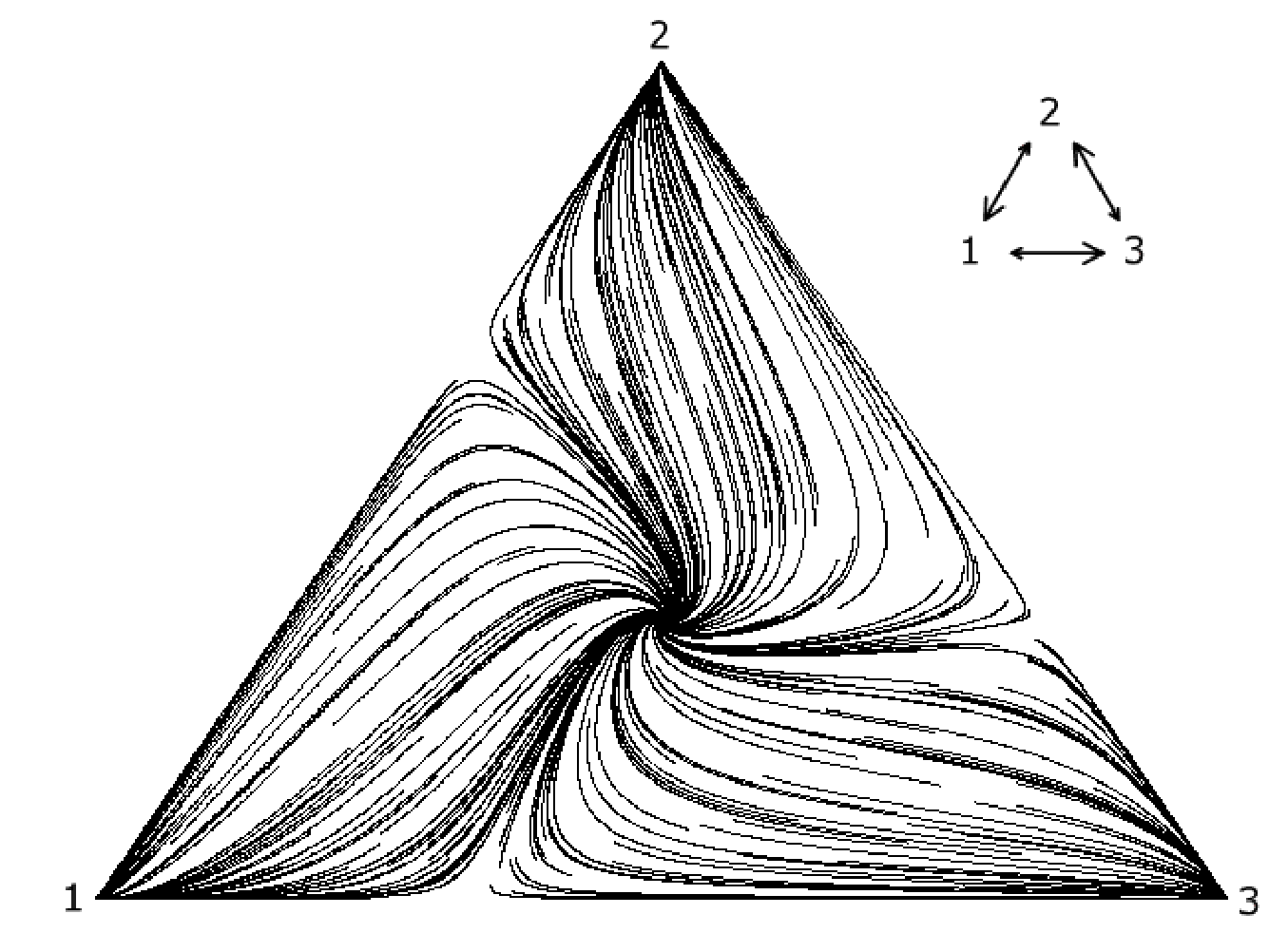}}}
 \caption{(a) Phase portrait  for the usual Sznajd Model 
(everybody convinces everyone with equal probabilities), in a mean
field approach:
 there  are three stable fixed points (vertices), that  
correspond to absorbing
 states of consensus with opinions $1$, $2$ and $3$; three saddle points,
 in which two opinions coexist, 
 and an unstable node with the coexistence of all three opinions. (b) The
 scenario described in (a) does not change qualitatively 
as long as the convincing
 capability between any two opinions is different from zero. In this picture 
$p_{1 \rightarrow 2}= 
 p_{2\rightarrow 3} = p_{3 \rightarrow 1} = 0.5$ and $p_{2 \rightarrow
 1} = p_{3 \rightarrow 2} = p_{1 \rightarrow 3} = 1$. The insets
resume the interacting rules; the size of the head of an arrow
indicates the strength of the convincing power in that direction.}
\label{usual}
\end{figure}

One can show that, if 
$p_{\sigma \rightarrow \sigma'} \neq 0\; \,\forall\, \;\sigma \neq \sigma '$,
(the usual SM lies in this class) the
flux has an unstable node, where all three opinions coexist; three 
saddle points, in which two opinions coexist; and three stable nodes,
representing consensus states (see appendix). Therefore, as
far as the convincing power among all different opinions is 
non-zero, the flux is qualitatively the same and the model will
always evolve to an absorbing state of consensus. However, the basins
of attraction change, and the same initial condition may belong to
different basins if the convincing capabilities change 
(see figure \ref{usual}). 


But what happens when  two opinions $\sigma$ and $\sigma'$ do not interact? 
This problem, for 
$p_{\sigma \rightarrow \sigma'} = p_{\sigma' \rightarrow \sigma} = 0$ if
$|\sigma - \sigma'|>1$, has already been studied in detail, both numerically 
\cite{Stauffer-2001} (square lattice) and in a mean-field approximation 
\cite{Schulze-2004}, with different conclusions. 
Stauffer showed that, in a square lattice, 
the stationary state is \emph{almost always} an absorbing state of
consensus in opinion 2. However, Schulze simulated the same model in
a complete graph - what corresponds to a mean-field approach - 
and only in 50\% of the simulations the model evolved to the consensus 
state with opinion 2, found by Stauffer; in the other cases, he observed 
a steady-state with coexistence of opinions 1 and 3.  
Our analytical approach and generalized model allow us to 
to understand why.

In order to understand why the model behaves differently in the mean-field and
in the square lattice, we first note that the BA network behaves in
the same way as the square lattice, almost always reaching consensus 
for opinion 2. One would expect the BA network to behave approximately 
like the mean field, as they both
have \emph{small world} properties, unlike the square lattice. 
So whatever process causes the lattice to always reach consensus 
must also be present in the BA network.

\begin{figure}[htbp!]
\subfigure[\label{grafico-B}]
{\centerline{\includegraphics[width=0.6\textwidth]{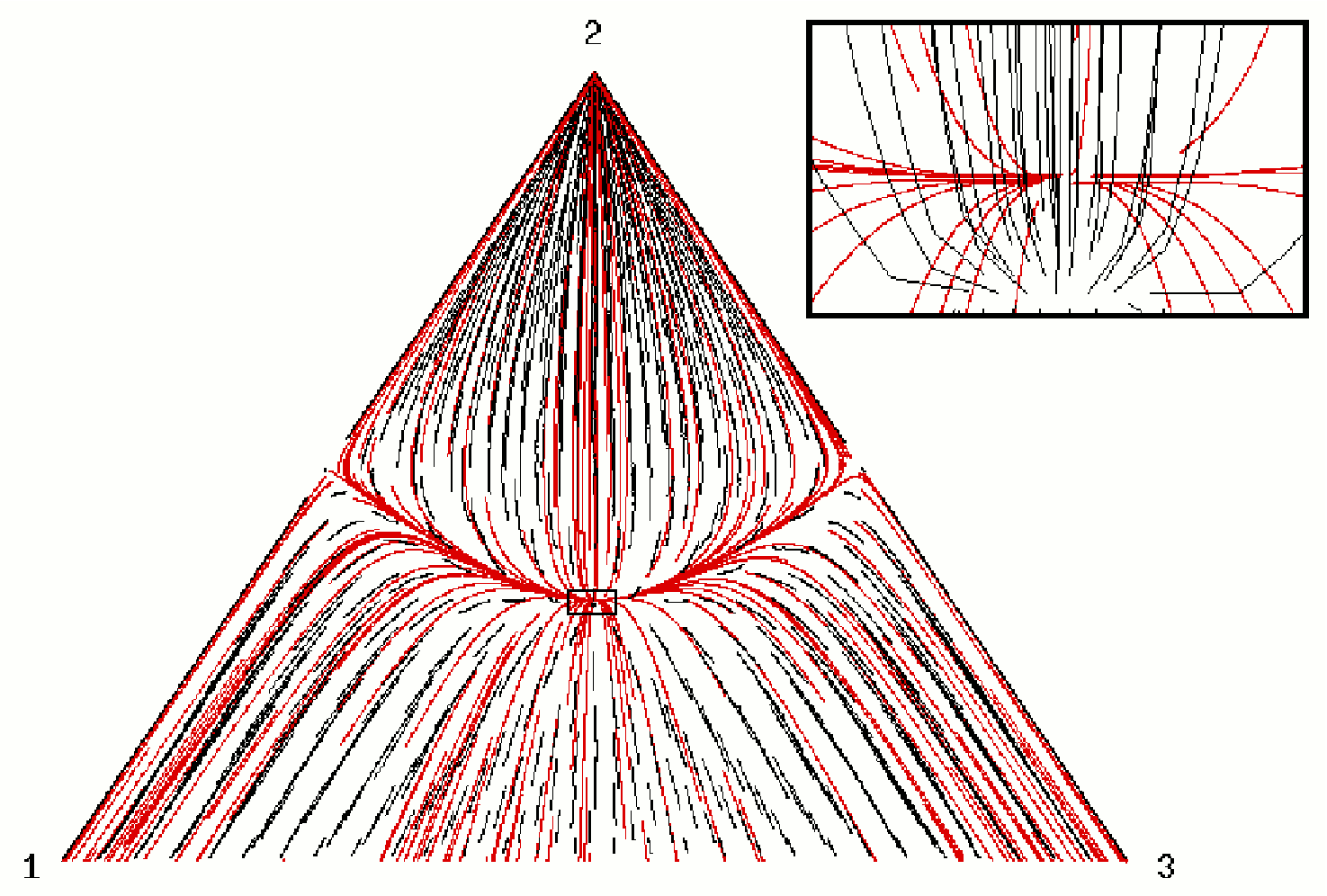}}}
\subfigure[\label{grafico-L}]
{\centerline{\includegraphics[width = 0.6\textwidth]{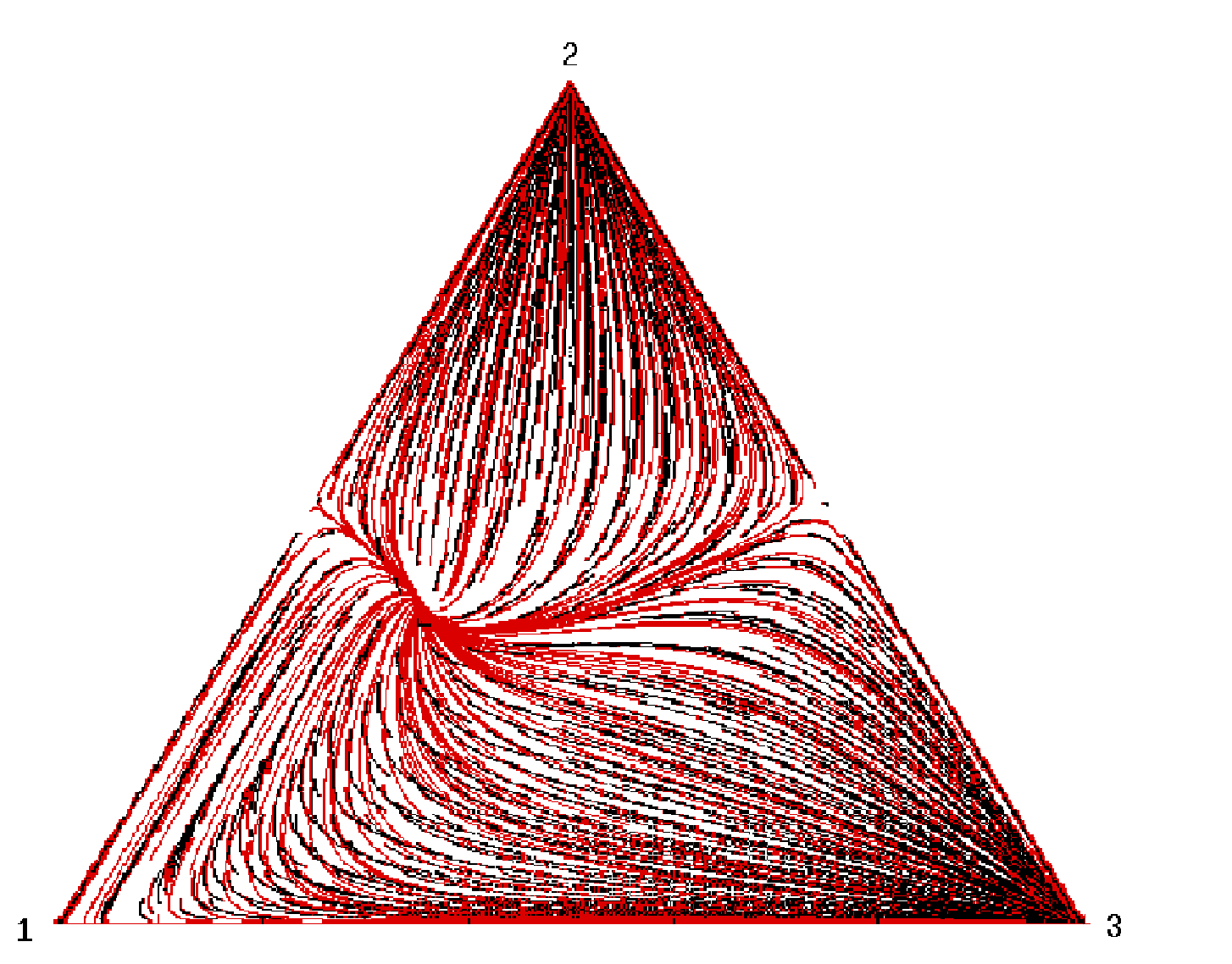}}}
\caption{Comparison between mean-field trajectories 
(in gray) and time evolution of the model in a BA network
(in black), for the case of three opinions and different combinations
of $p_{\sigma \rightarrow \sigma'}$. In (a) we have $\epsilon = 1$ 
(usual bounded confidence). The
only difference is a slight change in the position of the line
that separates the two basins of attraction (see detail of
central region in (a)).  In (b), $p_{1\rightarrow 2} = p_{1\rightarrow 3} = 
p_{2\rightarrow 1} = p_{2\rightarrow 3} = p_{3\rightarrow 2} = 1$ 
but $p_{3\rightarrow 1} = 0$, \emph{i.e.} sites with opinion 1 are unable
of convincing sites with opinion 3. Note that, in both cases, there is no 
qualitative change, (actually little quantitative changes) between the 
phase portraits of the two networks.}
\label{SMcomBC}
\end{figure}

To compare the results for the mean-field and the BA network 
we integrated numerically the equations for the mean-field, to get a
phase space portrait of the dynamics. Then, we built an `equivalent'
portrait for the stochastic model in a BA network, in the following way: 
we evolved the model from an initial condition chosen at random, 
but with specific expected values of $\eta_1, \eta_2$ and $\eta_3$, 
averaging over many simulations. Finally, we plotted the resulting 
\emph{trajectories}, together with the mean-field results 
(see figure \ref{SMcomBC}).


The picture shows that in both cases there are basins of attraction for two
kinds of solutions: consensus in opinion 2 or coexistence of
opinions 1 and 3. If the initial opinions are drawn at random, with
equal probability among opinions 1, 2 and 3, (as done in 
\cite{Stauffer-2001, Schulze-2004}),
 the initial condition will approximately lay in a circle of 
radius proportional
 to $\nicefrac{1}{\sqrt{N}}$, where $N$ is the number of sites, 
centered in the point $\eta_1 = \eta_2 = \eta_3 = \nicefrac{1}{3}$. 

In the mean-field scenario, this special point is located on the
border of the two basins of attraction. As a consequence, no matter 
how large is $N$ and, consequently, 
how small is the neighborhood around the point in which the 
initial conditions lay, half of its area will be in one basin of 
attraction and half in the other one.

On the other hand, for the stochastic model this point is, 
although close to the border, \emph{inside} the consensus basin 
of attraction. So the coexistence state can only be achieved for 
small values of $N$ (small lattices), for which the fluctuations 
in the initial condition are bigger.
 
For different choices of $p_{ \sigma \rightarrow \sigma'}$ the qualitative
behavior (fixed points, and basins of attraction) in phase space is only 
influenced by which of these probabilities are 0 and which are non-zero.
When a limit $p_{ \sigma \rightarrow \sigma'}\rightarrow 0$ is taken, typically
there will be some fixed points that collapse to already existing fixed points
where fewer opinions coexist. In the example (b) of figure
\ref{SMcomBC} the saddle point between $P_1$ and $P_3$ 
(coexistence of 2 opinions) collapses to the node in $P_1$
(only 1 opinion) that becomes a saddle.

For all the possibilities where $p_{ \sigma \rightarrow \sigma'}$ 
is either 0 or 1,  the mean-field approach is able to capture the whole  
qualitative behavior of the lattice model (see figure 
\ref{grafico-L} for instance), even though the \emph{trajectories}
representing the time evolution of the model in a lattice cross each other,
what is possible since it is not a flux. This is an assymetric case,
for which opinion 1(2) convinces 2(1), opinion 2(3) convinces 3(2),
but only opinion 3 is able to change opinion 1 ($p_{3 \rightarrow 1}=0$).

\begin{figure}[htbp]
\subfigure[\label{serie_quadrada}]
{\centerline{\includegraphics[width=0.3\textwidth]{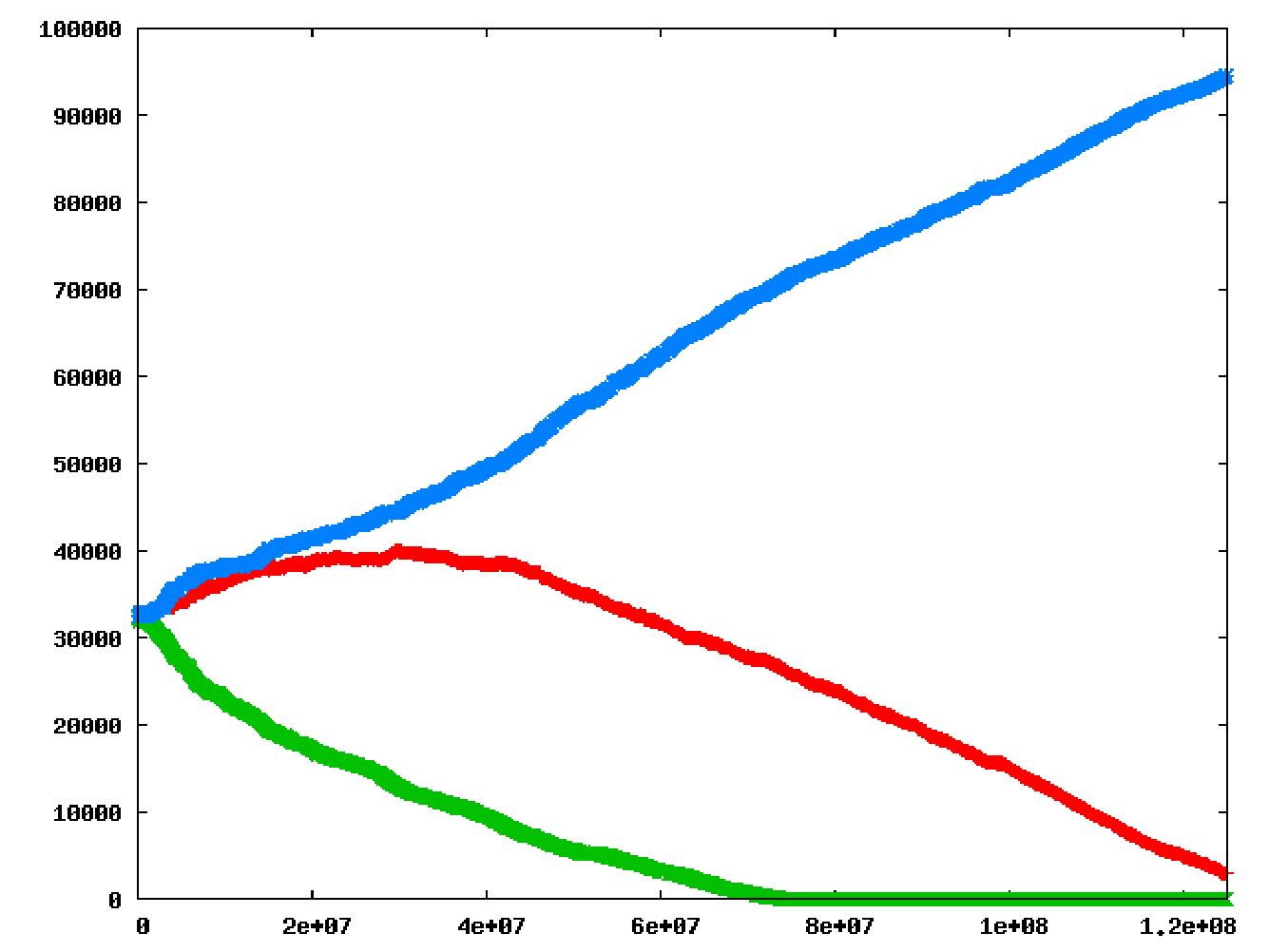}}}
\subfigure[\label{serie_BA}]
{\centerline{\includegraphics[width=0.3\textwidth]{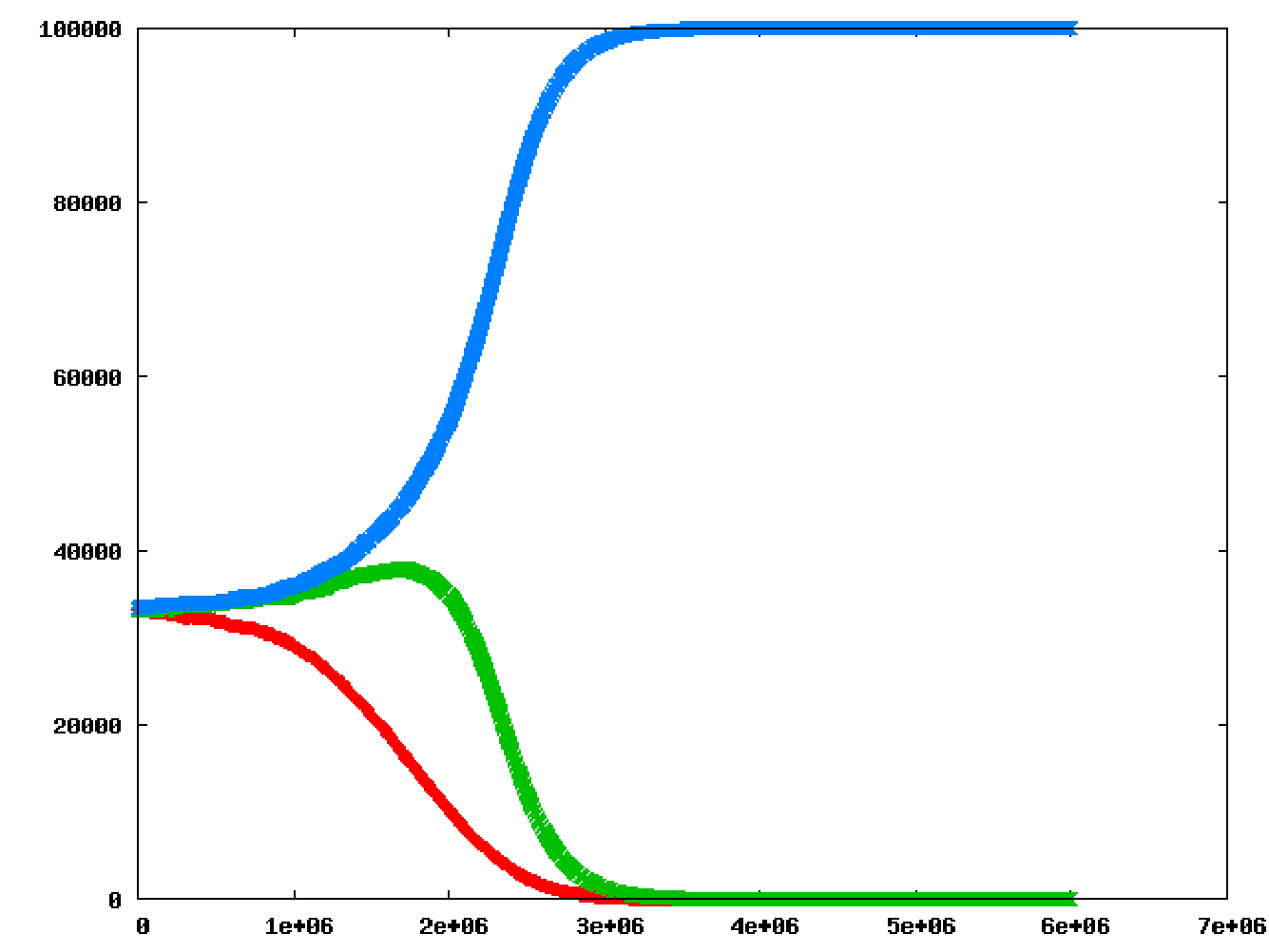}}}
\subfigure[\label{serie_CM}]
{\centerline{\includegraphics[width = 0.3\textwidth]{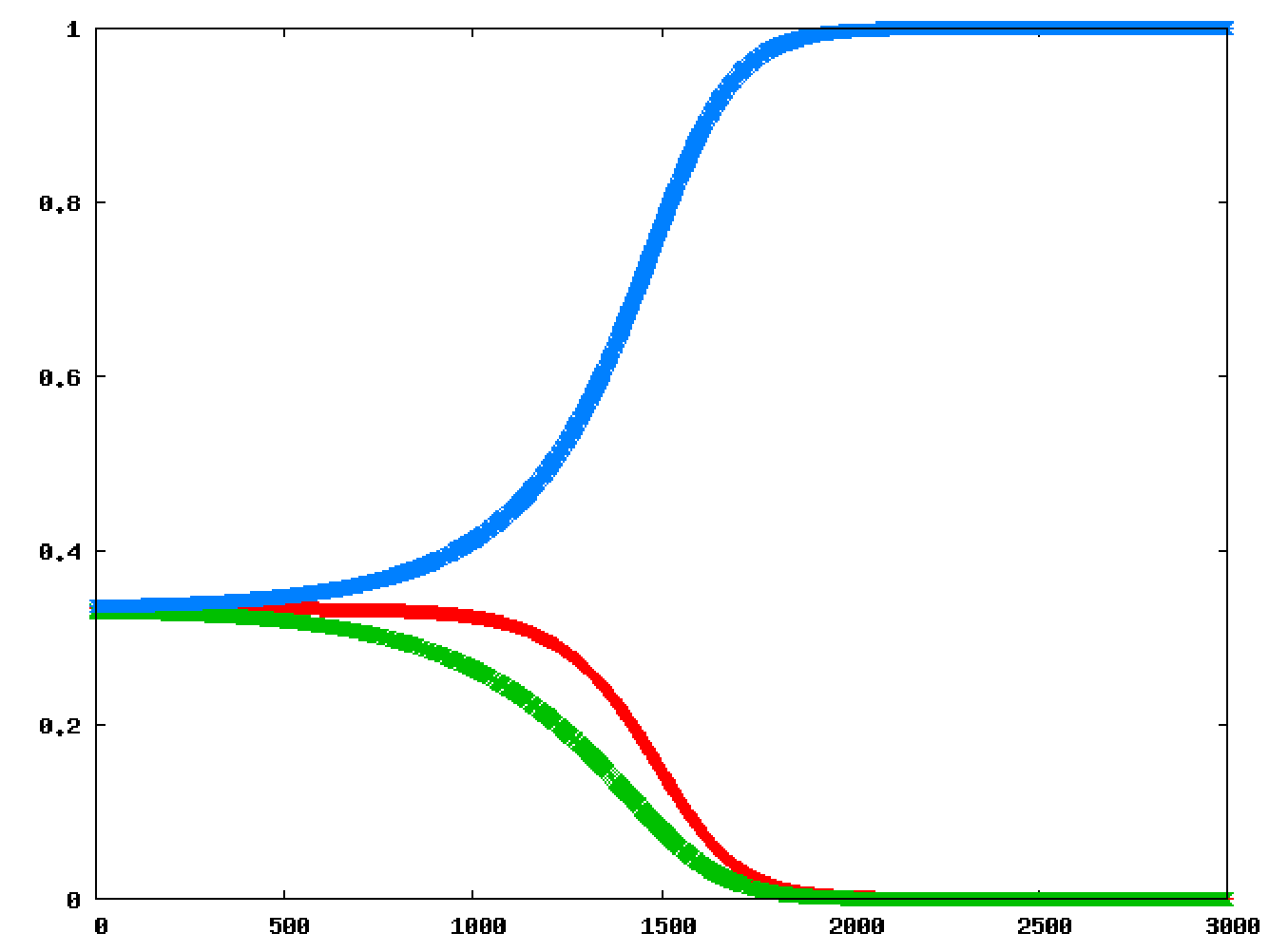}}}
\caption{Comparison between the time series of the Sznajd model
in different networks and in the absence of bounded confidence. Each color
 (gray-scale) represents a different opinion. The horizontal axis is time
 and the vertical one is the number (proportion for the mean field) of voters.
Graph (a) is for a square lattice (approximately $10^5$ sites), (b) is
for a BA network ($10^6$ sites and $m=5$) and (c) is for the mean-field. 
We can see the resemblance of the mean field and the BA network time series.}
\label{series_temporais}
\end{figure}

If we study the time evolution of the average number of votes of
each candidate, in the three situations studied (square lattice,
Barab\'asi-Albert network and mean-field), we see that the mean-filed is a
much better approximation for the BA case (see figure
\ref{series_temporais}).


\subsection{Scenario with four opinions}

\begin{figure}[htbp]
\centering
\includegraphics[width=0.4\textwidth]{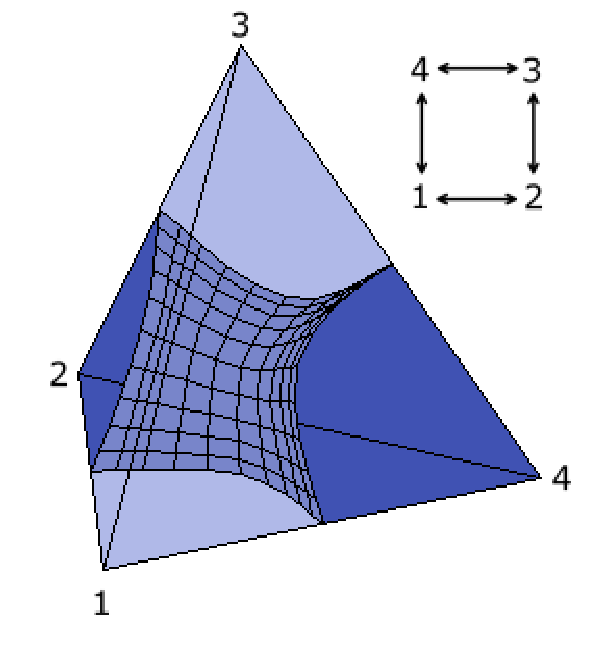}
\caption{(Color online) Boundary between the 2 basins of 
attraction for the four opinion model with $\epsilon = 1$; different
basins are in different gray (blue) tones.}
\label{grafico-fr}
\end{figure}

In the case of four opinions, $\eta_1 + \eta_2 + \eta_3 + \eta_4= 1$,
 and the flux is restricted to a tetrahedron. With usual bounded
 confidence rules, 
$p_{\sigma \rightarrow \sigma'} = p_{\sigma' \rightarrow \sigma} = 0$ if
$|\sigma - \sigma'|>1$ and $p_{\sigma\rightarrow\sigma'} = 1$ otherwise. 
If we add an interaction between opinions 1 and 4 
($p_{1\rightarrow 4} = p_{4\rightarrow 1} = 1$), each one of the 
tetrahedron's faces reproduces the 3 opinion scenario described in
the previous section. By continuity arguments, we can guess that in
this case, there are two distinct basins of attraction, shown in
figure \ref{grafico-fr}. The internal surface isolates completely
region I, that includes the edge $ 2 \Leftrightarrow 4$ from 
region II, that includes edge  $1 \Leftrightarrow 3$. 
As all points of both these edges are fixed points,
 there are two possible absorbing states with coexistence of opinions
 (opinions 2 and 4 or opinions 1 and 3), regions I and II are therefore
 the basins of attraction of these states. The fixed point with coexistence of 
 four opinions (that lies in the surface between I and II) is
 unstable, the ones with  three opinions are saddles 
(the edges $1\Leftrightarrow 4, 2\Leftrightarrow 3, 1\Leftrightarrow 2 $ 
 and $4\Leftrightarrow 3$ are unstable manifolds) and although 
the consensus states are in  the stable edges 
($1\Leftrightarrow 3$ and $2\Leftrightarrow 4$) they are unattainable.


\section{Conclusions}

In summary, we propose a new version of the Sznajd Model, generalizing
the bounded confidence rule. We solve the model in a mean-field approach, 
for a quite general case, discussing some aspects of the dynamics. We
showed that  the qualitative behavior of trajectories in the mean-field 
approach can be reduced to the study of the cases 
$p_{\sigma\rightarrow\sigma'} = 0$ and $p_{\sigma\rightarrow\sigma'} =
1$ for each one of the possible pairs of opinions
$\sigma$ and $\sigma'$.
Also, as long as every opinion interacts with all the others, the only 
possible absorbing state is consensus. 

For the special cases of three and four opinions, that had already been
studied in the literature, we were able to find
a nice way of representing the whole phase space and 
drew the detailed phase portrait, both in a mean-field approach 
and for a Barab\'asi-Albert network simulation (in which case we
developed a method to draw the stochastic \emph{trajectories}). 
In both cases the results
are qualitatively the same (in fact, they are remarkably alike), 
with two distinct basins of attractions:
one for an absorbing state of consensus in opinion 2, and another for
an absorbing state with coexistence of opinions 1 and 3. The only
difference was in the position of the line that separates the two
basins of attraction.
This picture enabled us to understand  why 
in \cite{Schulze-2004} (mean-field), 
the model ended up in an absorbing state
of consensus only in 50\% of the cases, while in the numerical
simulations on a BA network \cite{Stauffer-2001} 
it \emph{almost always} ended up in this state.

 Also, regarding the whole time evolution for the average number of
 electors, with opinion $\sigma$, we
were able to derive the following conclusions: (a) for three
opinions, a mean-field approach is  able to reproduce all the main
properties of the model when a complex network (usually a network with
small world properties) is employed to describe the relationship among the
electors; however, the same does not happen when the model is simulated on a
square lattice. 
(b) The existence of  any restriction
in the convincing power of agents, with at least two opinions that do
not interact one with the other may lead both in a mean-field
approach and in the BA network simulation to two classes of absorbing
states: consensus in one of the opinions that interacts with all
the others or coexistence of two (or more) opinions that do not
interact. The only difference between the 2 networks is in the
position of the basins of attraction, which means that the initial 
configuration is very important to define the asymptotic
behavior, and in order to understand such models, the
whole phase space must be taken into account. In particular, the
`natural' initial condition with a uniform distribution of opinions among
the voters may lay in different basins of attraction, if different
networks (or mean field) are employed. 

The new generalized model introduced by us put the
original Sznajd model and a variety 
of \emph{bounded confidence} versions together in a single model, and
the dynamical system's approach employed in its analysis 
 allowed to actually understand the whole model and to what
 extent the asymmetries in the way each opinion convinces the others
 can change qualitatively the behavior of the system.
 We believe that such approach, that is quite
general, can easily be adapted to unveil new features or draw 
unifying pictures of other similar models.

\bibliography{Timpanaro_arxiv09}

\section{Appendix}

\label{apendice_3op}

In the fixed point we have

\begin{subequations}\label{apendice1}
\begin{align}
& \eta_1+\eta_2+\eta_3 = 1\label{ap1-a}\\
& \eta_1^2\eta_2p_{2\rightarrow 1}-\eta_1\eta_2^2p_{1\rightarrow 2} + 
\eta_1^2\eta_3 p_{3\rightarrow 1}-\eta_1\eta_3^2 p_{1\rightarrow 3} 
= 0 \label{ap1-b}\\
& \eta_2^2\eta_1p_{1\rightarrow 2}-\eta_1^2\eta_2p_{2\rightarrow 1} + 
\eta_2^2\eta_3 p_{3\rightarrow 2}-\eta_2\eta_3^2 p_{2\rightarrow 3} 
= 0\label{ap1-c}.
\end{align}
\end{subequations}

From this set of equations its is trivial to show that, if only one of the
$\eta_{\sigma} \neq 0$, we have a stable fixed point in one of the
vertices;
If $\eta_{\sigma} \neq 0$ for all values of  $\sigma$, 
we can define $\mu_1 = \nicefrac{\eta_1}{\eta_3}$  and  
$\mu_2=\nicefrac{\eta_2}{\eta_3}$; equations \eqref{apendice1} can then
be written as  

\begin{subequations}\label{apendice2}
\begin{align}
\mu_1\mu_2p_{2\rightarrow 1} - \mu_2^2p_{1\rightarrow 2} + 
\mu_1p_{3\rightarrow 1} - p_{1\rightarrow 3}  = 0 \label{ap2-a}\\
\mu_1\mu_2p_{1\rightarrow 2} - \mu_1^2p_{2\rightarrow 1} + 
\mu_2p_{3\rightarrow 2} - p_{2\rightarrow 3}  = 0 \label{ap2-b}.
\end{align}
\end{subequations}

from \eqref{ap2-b}  we get

\[
\mu_2 = \frac{\mu_1^2p_{2\rightarrow 1} + 
p_{2\rightarrow 3}}{\mu_1p_{1\rightarrow 2} + p_{3\rightarrow 2}},
\]

and \eqref{ap2-a} becomes

\begin{equation}
\begin{split}
\mu_1p_{2\rightarrow 1}(\mu_1^2p_{2\rightarrow 1} + 
p_{2\rightarrow 3})(\mu_1p_{1\rightarrow 2} + p_{3\rightarrow 2}) - 
p_{1\rightarrow 2}(\mu_1^2p_{2\rightarrow 1} + p_{2\rightarrow 3})^2 
+ & \\
 +(\mu_1p_{3\rightarrow 1}-p_{1\rightarrow 3})(\mu_1p_{1\rightarrow 2}
 + p_{3\rightarrow 2})^2 & = 0,
\end{split}
\end{equation}

\noindent that is an ordinary polynomial of third order in $\mu_1$:

\begin{equation}
\begin{split}
f(\mu_1)  = &\, (p_{2 \rightarrow 1}^2p_{3 \rightarrow 2} + 
p_{3 \rightarrow 1}p_{1 \rightarrow 2}^2)\mu_1^3 - 
p_{1 \rightarrow 2}(p_{2 \rightarrow 1}p_{2 \rightarrow 3} - 
2p_{3 \rightarrow 1}p_{3 \rightarrow 2} + \\
& + p_{1 \rightarrow 2}p_{1 \rightarrow 3}) \mu_1^2 + 
p_{3 \rightarrow 2}(p_{2 \rightarrow 1}p_{2 \rightarrow 3} - 
2p_{1 \rightarrow 2}p_{1 \rightarrow 3} + 
p_{3 \rightarrow 1}p_{3 \rightarrow 2})\mu_1 - \\
& - (p_{1 \rightarrow 2}p_{2 \rightarrow 3}^2 + 
p_{1 \rightarrow 3}p_{3 \rightarrow 2}^2) = A\mu_1^3 + B\mu_1^2 + C\mu_1 + D = 0.
\end{split}
\end{equation}

\noindent The real positive roots of this polynomial corresponds 
to fixed points in which all three opinions coexist; 
if $p_{\sigma\rightarrow\sigma'}> 0$, $A>0$ and $D<0$, 
there is 1 or 3 positive roots. 

Suppose,
by absurd, that there were three real positive roots: we 
would then have  $B<0$, $C>0$ and the
discriminant  $\Delta = 4AC^3 + 4B^3D - B^2C^2 + 27A^2D^2 - 18ABCD \leq 0$; 

\noindent defining  

\[
\alpha = \frac{C}{3(AD^2)^{\frac{1}{3}}} \qquad \mathrm{and}\qquad  
\beta = \frac{B}{3(A^2D)^{\frac{1}{3}}},
\]

\noindent we note that, if $A,C>0$ and 
$B,D<0,\,$ we  have  $\alpha ,\beta >0$. 
Besides, 
$\,C^3 = 27\alpha^3AD^2\,$,$\,B^3  = 27\beta^3A^2D \,$ and 
$BC  =9AD\alpha\beta$,

\noindent so that
\[
\Delta = 108\alpha^3A^2D^2 + 108\beta^3A^2D^2 - 
81\alpha^2\beta^2A^2D^2 + 27A^2D^2 - 162\alpha\beta A^2D^2
\]
\noindent and 
\[
\delta = \frac{\Delta}{27A^2D^2} = 4\alpha^3 + 
4\beta^3 + 1 - 3\alpha^2\beta^2 - 6\alpha\beta \leq 0,
\]

\noindent that is

\[
\delta(\alpha ,\beta) = E(\beta)\alpha^3 + F(\beta)\alpha^2 + 
G(\beta)\alpha + H(\beta),
\]

\noindent where $E = 4 , F = -3\beta^2 , G = -6\beta$ and $H=4\beta^3 +
1$ (note that $H>0$). 
So, if we fix $\beta$, $\delta(\alpha)$ has a negative root. 
But $E>0$, and, unless $\delta$ has a positive root, we would have 
$\delta > 0$ for  $\alpha>0$; 
However, this is only possible if $\delta$ has three real roots, what
is equivalent to  

\[
4EG^3 + 4F^3H - F^2G^2 + 27E^2H^2 - 18EFGH \leq 0 
\]
\noindent or
\[
\begin{split}
& - 3456\beta^3 - 108\beta^6(4\beta^3 + 1) - 324\beta^6 
+ 432(16\beta^6 + 8\beta^3 + 1) - 1296\beta^3(4\beta^3 + 1)= \\
= \,& -432(\beta^3 - 1)^3 \leq 0 \Rightarrow \beta \geq 1.
\end{split}
\]

\noindent But because $\delta(\alpha , \beta) = \delta(\beta ,
\alpha)$, we have 
$\alpha \geq 1 \Rightarrow \alpha\beta\geq 1 \Rightarrow BC\leq 9AD$. 
However, we also have

\[
\begin{split}
BC-9AD = & \,\,8p_{2\rightarrow 1}^2p_{1\rightarrow 2}p_{3\rightarrow 2}
p_{2\rightarrow 3}^2 + 9p_{2\rightarrow 1}^2p_{3\rightarrow 2}^3
p_{1\rightarrow 3} + 9p_{3\rightarrow 1}p_{1\rightarrow
  2}^3p_{2\rightarrow 3}^2 + \\
& + 4p_{3\rightarrow 1}p_{1\rightarrow 2}^2p_{3\rightarrow 2}^2
p_{1\rightarrow 3} + p_{2\rightarrow 1}p_{1\rightarrow 2}^2
p_{3\rightarrow 2}p_{1\rightarrow 3}p_{2\rightarrow 3} + \\
& + p_{2\rightarrow 1}p_{3\rightarrow 1}p_{1\rightarrow
  2}p_{3\rightarrow 2}^2
p_{2\rightarrow 3} + 2p_{3\rightarrow 1}^2p_{1\rightarrow 2}
p_{3\rightarrow 2}^3 + 2p_{1\rightarrow 2}^3p_{3\rightarrow 2}
p_{1\rightarrow 3}^2 > 0, 
\end{split}
\]

\noindent what is a contradiction! So, there is one and only one positive
root, and only one (unstable) fixed point inside the triangle. The  
corresponding values of $\eta$  can be obtained from \eqref{ap1-a}:
\begin{equation*}
\eta_1  = \frac{\mu_1}{\mu_1+\mu_2+1}; \qquad 
\eta_2  = \frac{\mu_2}{\mu_1+\mu_2+1} \qquad \mbox{and} \qquad 
\eta_3  = \frac{1}{\mu_1+\mu_2+1}.
\end{equation*}


Finally, it is also easy to see that, if only one of the $\eta$ are equal to
zero in \eqref{apendice1}, we have three other (stable) fixed points: 
\begin{align*}
\eta_1 & = 0, \qquad \eta_{2(3)}  =
\frac{p_{2(3)\rightarrow 3(2)}}{p_{2(3)\rightarrow 3(2)}+ 
p_{3(2)\rightarrow 2(3)}};\\
\eta_2 & = 0, \qquad \eta_{3(1)}  =
\frac{p_{3(1)\rightarrow 1(3)}}{p_{3(1)\rightarrow 1(3)}+ 
p_{1(3)\rightarrow 3(1)}}; \qquad \mbox{and} \\
\eta_3 & = 0, \qquad \eta_{1(2)}  =
\frac{p_{1(2)\rightarrow 2(1)}}{p_{1(2)\rightarrow 2(1)}+p_{2(1)\rightarrow 1(2)}}.
\end{align*}

\noindent By continuity reasons there must be saddle points within the unstable
manifolds along the sides.

\end{document}